\documentclass[10pt,letterpaper]{article}
\usepackage[top=0.85in,left=2.75in,footskip=0.75in]{geometry}

\usepackage{amsmath,amssymb}

\usepackage{changepage}

\usepackage[utf8x]{inputenc} 
\usepackage[T1]{fontenc} 

\usepackage{textcomp,marvosym}

\usepackage{cite}

\usepackage{nameref,hyperref,url}
\makeatletter
\g@addto@macro{\UrlBreaks}{\UrlOrds}
\makeatother

\usepackage[right]{lineno}

\usepackage{microtype}
\DisableLigatures[f]{encoding = *, family = * }

\usepackage[table]{xcolor}

\usepackage{array}

\newcolumntype{+}{!{\vrule width 2pt}}

\newlength\savedwidth

\raggedright
\usepackage[aboveskip=1pt,labelfont=bf,labelsep=period,justification=raggedright,singlelinecheck=off]{caption}

\makeatletter
\renewcommand{\@biblabel}[1]{\quad#1.}
\makeatother

\usepackage{lastpage,fancyhdr,graphicx}
\usepackage{epstopdf}
\fancyheadoffset[L]{2.25in}
\fancyfootoffset[L]{2.25in}
\usepackage{booktabs}
\usepackage{color}

\begin{document}
\vspace*{0.2in}

\begin{flushleft}
{\Large
\textbf\newline{User perspectives on critical factors for collaborative playlists} 
}
\newline
\\
So Yeon Park\textsuperscript{1,2,*},
Blair Kaneshiro\textsuperscript{2}
\\
\bigskip
\textbf{1} Center for Design Research, Stanford University, Stanford, CA, USA
\\
\textbf{2} Center for Computer Research in Music and Acoustics, Stanford University, Stanford, CA, USA
\\
\bigskip

%
%


\textcurrency Current Address: Center for Design Research, Stanford University, 424 Panama Mall, Stanford, CA USA 94305



* syjpark@stanford.edu

\end{flushleft}
\section*{Abstract}
Collaborative playlists (CP) enable listeners to curate music together, translating long-standing social practices around music consumption into the age of streaming. Yet despite their role in connecting people through music, we lack an understanding of factors that are critical to CPs and their enjoyment.
To understand what users consider important to CPs and their usage, we investigated aspects that are perceived to be most useful and lacking in today's CP implementations. We conducted a survey to collect open-ended text responses from real-world CP users. Using thematic analysis, we derived the Codebook of Critical CP Factors, which comprises eight aspects. We gained insights into which aspects are particularly useful, and which are absent and desired by current CP users. From these findings we propose design implications to inform further design of CP functionalities and platforms, and highlight potential benefits and challenges related to their adoption in current music services.



\section*{Introduction}\label{sec:introduction}
Selecting and listening to music together are long-standing social activities. One such example, music co-curation, has a rich history spanning multiple use cases over the past century, from technologies and artifacts predating digital music collections (e.g., jukeboxes, mix tapes) to today's usage of music streaming platforms. One way in which users can curate music together on streaming platforms is by co-editing a collaborative playlist (CP), which is ``a list of songs that multiple users have created using a digital platform''~\cite{park2019tunes}. CP functionalities have been available on major commercial streaming platforms for over a decade, enabling users to socially curate and consume music in a manner similar to personal music curation. 

However, relatively little attention has been paid to CPs. In terms of platform implementation, only a few major streaming services such as Spotify, Deezer, and YouTube have implemented CP functionalities, and, to the best of our knowledge, these implementations are essentially personal playlists with co-editing functionalities allowing multiple users to add, delete, or reorder tracks~\cite{dummies}. 
Moreover, while the study of present-day music consumption is an established sub-field of music information retrieval (MIR)~\cite{lee2013towardAnUnderstanding}, most user research relating to music streaming platforms---e.g., characterizing tastes~\cite{ferwerda2017personality}, recommending songs~\cite{schedl2018current}, or facilitating search~\cite{hosey2019search}---focuses on individual usage, while studies of CPs and their usage are fewer in number~\cite{park2017anAnalysis,park2019tunes,park2021social,park2021lost}. Consequently, our understanding on how CPs are used, and how streaming platforms can best support users in the social curation setting, is relatively lacking.

In 2020, shutdowns related to the COVID-19 pandemic forced many human-to-human interactions to become virtual~\cite{roose2020liveOnline}. As a result, a number of music-related activities have been impacted, including performing together~\cite{randles2020performOnline}, taking lessons~\cite{kane2020lessonsOnline}, and attending concerts~\cite{okane2020concertsOnline}. CPs, too, have received added attention at this time~\cite{mihaly2020covidCP,fortherecord2020your,fortherecord2020how}---possibly contributing to recent design improvements. While the proliferation of virtual or socially distanced musical activities is currently a necessity, these trends may ultimately become long-term, fundamental changes to how we interact~\cite{politico2020longTerm}. Thus there is a need, now more than ever, for well-designed social platforms---including those created for musical activities.

In this study we extend the literature on CPs by focusing specifically on user perceptions of CPs and their usage. As our main research question, we ask,
\begin{center}\textit{What aspects of CPs and their experience are most important to users?}\end{center}
\noindent We address this question by unpacking free-text responses from $N=70$ CP users, who reported on (1) what they feel is most useful and (2) what they find missing from existing CP platforms. From a thematic analysis of these responses, we derived the Codebook of Critical CP Factors, comprising eight aspects of CPs and their usage. We then performed consensus coding to label the responses according to the categories of the codebook.
Based on the extent to which the responses collectively implicated---positively and negatively---each of the eight Codebook categories, and insights from related literature, we have derived design implications to inform future design of CP platforms. Taken together, this study extends the body of user research in music information retrieval---and human-computer interaction more broadly---by providing much-needed insights into digitally mediated social music curation, while also informing real-world platform design.

\section*{Related Works}\label{sec:relatedWorks}
We provide background on music curation from prior literature on general music collections as well as CPs. As CPs are one way in which users engage with one another through music, we also expand upon social music activities. Finally, we relate CPs to social music prototypes and collaborative platforms, as findings from these studies may inform design of CP platforms.

\subsection*{Music curation and playlists}
Music curations have existed in many forms, from LP collections to mixtapes. Today, collections of songs are often organized digitally as playlists~\cite{hagen2015playlist}. Playlists---generally assumed to be for personal use---can be created by users or provided by the streaming platform. They can be centered around specific themes or contexts, such as holidays~\cite{krause2018tisTheSeason} or a particular year~\cite{hagen2015playlist}, or accompany everyday activities such as working or exercising~\cite{hagen2015playlist,cunningham2017exploring,cunningham2019interactingWithPersonal}. Playlists can serve as a static record of music originally added, or be updated continuously~\cite{hagen2015playlist}. Their ease of access on streaming platforms enables users to curate music, share it publicly, and even gift music as digital mixtapes (\url{https://developer.spotify.com/documentation/general/guides/working-with-playlists/}).
Services such as 8tracks (\url{https://8tracks.com/duewets}), Playlists.net (\url{https://playlists.net/}), and Art of the Mix (\url{http://www.artofthemix.org/}) enable users to share (i.e., let others listen to but not co-curate) playlists online.

Social music curation and consumption are carried out on streaming platforms as well. Playlists can now be curated by and shared among users~\cite{hagen2015playlist,cunningham2019interactingWithPersonal}, with CP functionalities that enable multiple users to listen to or edit a playlist from multiple devices~\cite{dummies}.
Following a preliminary study exploring evolution, usage, and perceptions around CPs~\cite{park2017anAnalysis}, Park et al.\ introduced the CP Framework, which includes three purposes---Practical, Cognitive, and Social---motivating real-world users to engage with these shared playlists~\cite{park2019tunes}. Here, Practical purposes have to do with characteristics of the CP, the curation process, and consumption contexts; Cognitive purposes relate to learning and discovery about music in the CP or ones' collaborators; and Social purposes have to do with sharing the playlist, sharing music, or bonding and connecting with collaborators. More recently, Park \& Kaneshiro conducted a mixed-methods investigation in order to characterize successful CPs and their usage~\cite{park2021social}. Experimental research further underscores social effects on music curation: Bauer \& Ferwerda found that positive and negative judgments of simulated collaborators (i.e., bots) differentially affected participants' curation decisions~\cite{bauer2020conformityBehavior}, while Park \& Lee found that perceived ownership of CPs and their songs greatly influenced engagement~\cite{park2021lost}.

Even so, relatively little is known about CPs and their usage.
The bulk of innovation on today's music streaming platforms is geared toward personalization over socialization~\cite{schedl2018current}, and today's CPs embody essentially what we term a ``personal-plus'' implementation---that is, a personal playlist interface with co-editing capabilities. Yet the ideal CP design may be something quite different. 
CP user studies have provided tangible first insights into user needs~\cite{park2017anAnalysis,park2019tunes,park2021social}; however, the potential of experiential reports to inform ideal design specifications of CPs is limited by current affordances of the CP platform. Therefore, it is worth examining related fields of research as well, which may indirectly inform CP design. 

\subsection*{Social music activities}
Selecting and consuming music is a pleasurable activity, whether undertaken alone or with others ~\cite{laplante2011theUtilitarianAnd,brown2006sharing,cunningham2003ethnographic}. Curating music for others has evolved with technology over the past century, from dedicating songs to others over the radio~\cite{bradley2015callingArt} to the use of jukeboxes in social settings~\cite{frank2009futurehit} and the rise of mixtapes and CDs~\cite{voida2005listeningInPractices,cunningham2006moreOfAnArt}. As music collections migrated to digital formats such as iTunes and Napster in the early 2000s, studies of content sharing and curation over those platforms followed~\cite{voida2005listeningInPractices,brown2006sharing}. Other studies have investigated practices of music selection for shared in-person consumption---e.g., for road trips~\cite{cunningham2014socialMusicIn} and parties~\cite{cunningham2009exploringSocialMusic}, or in the home~\cite{leong2013revisitingSocialPractices}.

In terms of social practices on commercial streaming platforms, Spinelli and colleagues conducted focus groups to derive a codebook comprising nine social practices and 24 influences, from which they offer design implications for streaming services~\cite{spinelli2018users}. Social media, too, enables users to effortlessly share musical interests and links to the music itself~\cite{dewan2014socialMediaTraditional}, share sentiments about music~\cite{hubbles2017f}, and connect with artists~\cite{baym2012fansOrFriends} and other fans~\cite{santero2016nobody,chang2018the,park2021armed}. Since 2020, COVID-19 restrictions have forcibly reshaped the ways in which people consume music together. For example, livestreamed virtual concerts fulfill some (but not all) of the social needs of concertgoers~\cite{vandenberg2020lonelyRaver}, while Tim's Twitter Listening Party (\url{https://timstwitterlisteningparty.com/}) invites artists to live-tweet as they and their fans listen to one of their albums remotely yet synchronously; these events are thought to provide a new means for collective socialization and reminiscence around music~\cite{lee2020youKnowYoure}. 

\subsection*{Social music prototypes}
Field studies involving digital social music prototypes date back over 20 years. While these studies do not reflect real-world usage of widely used services, their insights highlight specific affordances and features that cannot currently be observed on commercial platforms and can potentially inform design implications for real-world CP platforms. 

An early example, MusicFX, selected music to broadcast in a gym using a group preference agent~\cite{mccarthy1998musicfx}. Subsequent proximity-based systems designed to broadcast group playlists include Flytrap~\cite{crossen2002flytrap} and Adaptive Radio~\cite{chao2005adaptiveRadio} systems; in these systems, music was selected for consumption in shared settings based upon listeners' positive and negative preferences, respectively, while Jukola enacted a voting system for music selection in a bar~\cite{ohara2004jukola} and Sound Pryer shared music among drivers in cars based on proximity~\cite{ostergren2004soundPryer}. 
Other systems were aimed toward mobile consumption. For instance, the tunA system enabled users to browse playlists, bookmark songs, and message with nearby users~\cite{bassoli2006tunaSocialisingMusic}, while Push!Music implemented a more tangible form of music sharing by permitting automatic copying or manual sharing of songs across users' devices~\cite{hakansson2007pushMusic}. Social Playlist, aiming to support already-established relationships among users, provided ``a shared playlist where members associate music from their personal library to their activities and locations''~\cite{liu2008social}. More recently, Kirk et al.\ introduced Pocketsong, which allowed users to observe what others were listening to, as well as ``gift'' snippets of songs to others~\cite{kirk2016understanding}. Finally, the MoodPic system from Lehtiniemi and colleagues enabled users to collaboratively curate music by associating songs to ``mood pictures''~\cite{lehtiniemi2013evaluating,lehtiniemi2017socially}.

Numerous valued (or requested, if not implemented) attributes of social music systems have emerged across these prototype studies. For instance, listening, sharing, and voting behaviors of others helped users discover music~\cite{ohara2004jukola,bassoli2006tunaSocialisingMusic,hakansson2007pushMusic}, learn about their friends~\cite{liu2008social}, and even discover others with similar tastes~\cite{bassoli2006tunaSocialisingMusic,lehtiniemi2017socially}. Such discoveries can lead to surprise and even serendipity, e.g., when encountering shared musical tastes~\cite{chao2005adaptiveRadio} or songs contributed by others~\cite{liu2008social,lehtiniemi2013evaluating}. In addition, social functionalities such as commenting, messaging, rating, and voting were usually viewed positively~\cite{ohara2004jukola,bassoli2006tunaSocialisingMusic,liu2008social,lehtiniemi2013evaluating,kirk2016understanding,lehtiniemi2017socially} and noted as ways to potentially strengthen existing social relationships~\cite{liu2008social} or form new ones~\cite{bassoli2006tunaSocialisingMusic}, clarify song selections~\cite{liu2008social}, and facilitate music discovery~\cite{lehtiniemi2017socially}. Finally, some users felt gratified knowing that others had listened to a song they contributed ~\cite{liu2008social,lehtiniemi2013evaluating}.

Prototype studies also highlight challenges and complications unique to social scenarios. Lehtiniemi \& Ojala reported varying, sometimes conflicting requests for collaborators' editing control of a playlist~\cite{lehtiniemi2013evaluating}; others noted that a collaborative system could produce ``simply an overflow of songs''~\cite{hakansson2007pushMusic} or bad songs in particular~\cite{liu2008social}. Despite positive aspects of social features noted above, some users worried that too much extramusical content in the platform could ``reduce the main role of the music in the service''~\cite{lehtiniemi2013evaluating}. Finally, while some users expressed an interest in knowing what others were listening to~\cite{hakansson2007pushMusic,lehtiniemi2013evaluating,kirk2016understanding}, others felt shy about adding songs (e.g., because they did not know the collaborator~\cite{lehtiniemi2013evaluating} or collaborators' tastes~\cite{kirk2016understanding}) and noted that they might change their listening behaviors, should those behaviors be made visible to others~\cite{kirk2016understanding}.

\subsection*{Other collaborative platforms}
While CPs differ from other types of collaborative platforms in critical ways~\cite{park2021social}, they may be broadly considered a form of collective content, defined by Olsson, in that they are ``digital media content that is regarded as \emph{commonly owned} as well as \emph{jointly created} and \emph{used}'', and ``both a consequence of collaborative activities with content and [...]\ a reason and motivator for collaborative activities to occur around content''~\cite{olsson2009understandingCollectiveContent}. Thus, other collaborative platforms may also provide insights relevant to user needs for CPs.

Numerous studies have investigated collaborations over digital platforms. Collaborative writing is prolifically studied in computer-supported cooperative work, and has led researchers to design for version control management~\cite{kraut1992task,baecker1995user}, annotations~\cite{baecker1995user,weng2004asynchronous}, and access methods~\cite{posner1992howPeopleWrite,neuwirth2001computer}. Services such as Google Docs provide functionalities that enable synchronous co-editing (\url{https://www.google.com/docs/about/}) multiple access levels (i.e., view, suggest, edit)~\cite{murphy2016how}, and an easily accessible version history through which users can view others' edits and revert to previous versions. Such functionalities can be beneficial to making contributions more visible, and thereby increase change awareness~\cite{tam2006framework}; however, they can also bring about social conflict~\cite{kittur2007he}. One study on collaborative authors has found that edits are made in socially conscious ways to mitigate such conflicts~\cite{birnholtz2012tracking}. Other features, such as the display of who is currently viewing or editing the document, promote ``workspace awareness'', an aspect of in-person collaboration which must be intentionally designed for in virtual settings~\cite{gutwin1996workspaceAwareness,gutwin1998effects}. Finally, studies of co-editing on Wikipedia have identified collaborative elements underlying article quality~\cite{kittur2008harnessingTheWisdom}, and have proposed user personas---such as ``zealots'' and ``Good Samaritans''---as a means of elucidating the nature and value of different collaboration styles~\cite{anthony2009reputationAndReliability}.

Online collaborations are also known to support social functions. Some online communities, such as special-interest groups, are formed with a social component in mind~\cite{olsson2009collectiveContentAs,olsson2009understandingCollectiveContent}; when a technology platform and the collaborative contributions enacted therein combine effectively, the ``Snowball effect'', described by Olsson and colleagues  as ``reciprocal activity that maintains or increases content-related and social interaction''~\cite{olsson2009collectiveContentAs}, can be achieved.

A number of known challenges are identified in past research. Collaborations can be hindered by access---for example, if not all collaborators use the same platform~\cite{posner1992howPeopleWrite}, or if a cloud-based system does not permit offline editing, as was previously the case with Google Docs~\cite{perron2011reviewOfCollaborative}. While collective content has the potential to enter a positive cycle of engagement~\cite{olsson2009collectiveContentAs},  ``social loafing''---the reduction in individuals' efforts when working as part of a group~\cite{latane1979many,williams1991social,karau1993social}---is known to be a general issue in online communities~\cite{rafaeli2004lurking,rafaeli2008online,kraut2012building}. As suggested in prior CP work~\cite{park2017anAnalysis,park2021lost}, uneven contributions and differing perceptions of ownership can lead to territoriality issues, which have also been noted around collaborative authoring in Wikipedia~\cite{halfaker2009jury,thom2009s} and co-curation in Pinterest~\cite{schiele2013possession}.

\section*{Methods}\label{sec:methods}
\subsection*{Ethics statement}
This study was approved by Stanford University's Institutional Review Board. All participants confirmed their eligibility, and indicated informed consent by agreeing to conditions detailed in an Information Sheet (Waiver of Consent), prior to participating in the study.

\subsection*{Survey and participants} 
We conducted a survey to better understand use cases around CPs from real-world users.
In order to ground participants, we first provided the following definition of CPs from~\cite{park2019tunes}: ``A list of songs that multiple users have created using a digital platform''. Then, the users answered various questions about their favorite CPs, which are not analyzed here. We then asked users about ``features'' to enable them to think in a more tangible way, and to enable us to derive aspects that are necessary for positive collaborative experiences---these questions are considered in the present analysis. The specific questions in the survey, to which CP users provided free-text responses, are as follows:

\vspace{\baselineskip}

\noindent \textbf{Q1}: What features of a collaborative playlist are most important or useful to you?\\
\noindent\textbf{Q2}: What are some shortcomings that you see in today's collaborative playlist platforms? What are some features that you believe would enhance the collaboration between you and your playlist collaborators?

\vspace{\baselineskip}

While open-ended questions are known to introduce challenges around the quality of responses and added analysis effort~\cite{pietsch2018topicModelingFor,reja2003openEndedVs,ladonna2018whyOpenEnded,zull2016openEndedQuestions,miller2014openEndedSurvey,jackson2002conceptMappingAs}, 
they also allow respondents to elaborate on their thoughts~\cite{zull2016openEndedQuestions} and express ideas that researchers may not have thought of as a priori response options~\cite{reja2003openEndedVs,jackson2002conceptMappingAs,pietsch2018topicModelingFor}. 
Moreover, open-ended responses are deemed useful at initial stages of research to ``[classify] the structure of a problem in all its details''~\cite{lazarsfeld1944controversyOverDetailed}, and are recommended ``if it is not yet possible to clearly delimit the subject of inquiry, or if one expects new topics to emerge''~\cite{zull2016openEndedQuestions}. 
As the present study is to the best of our knowledge the first exploration of user needs around CPs, we thus deemed the open-ended format to be appropriate.

\subsection*{Analysis of survey data}
Seeking to identify aspects of CPs that emerge from the data---including those that could be overlooked in the application of a pre-existing theoretical framework---we took an inductive approach to data analysis~\cite{thomas2006generalInductiveApproach,bryman2016socialResearchMethods}. Recent user research in MIR using similar such approaches has successfully characterized users of streaming services~\cite{lee2015understandingUsersPersonas}, user comments on SoundCloud~\cite{hubbles2017f}, social practices surrounding streaming services~\cite{spinelli2018users}, and purposes for engaging with CPs~\cite{park2019tunes}. 

We analyzed responses across the two questions, as they collectively implicated shared aspects of CPs and their usage. First, we aggregated all of the responses, and used thematic analysis to manually group individual ideas expressed therein~\cite{bryman2016socialResearchMethods}. A total of 18 groups emerged from this analysis. Next, when possible, emergent categories were grouped further under higher-level themes (e.g., ``Share'', ``Recommend'', ``Connect'', and ``Heart behind item'' all relate to social activities). The final set of high-level groupings formed the eight categories of the Codebook of Critical CP Factors (Table~\ref{tab:Codebook}). 
We subsequently documented the codebook with descriptions to aid in the coding process~\cite{macqueen1998codebookDevelopmentFor}. 
Once the codebook was finalized, we performed consensus coding over the full set of responses. Each free-text response could receive multiple codings; responses were coded at the category level as well as the sub-category level, when sub-categories existed. We two authors each independently coded all of the text responses, which yielded reasonable inter-rater reliabilities (Krippendorff’s alpha: 0.96 for Q1, 0.92 for Q2). All discrepancies between coders were resolved through discussion.

With few exceptions, responses to Q1 implicated positive perceptions, which we report as ``Useful/Important'' aspects of CPs and their usage; responses to Q2, reported as shortcomings or missing features, are framed as ``Lacking/Desired'' aspects. We report percentages of responses implicating each of the eight aspects (and their sub-categories, when applicable) as well as illustrative quotes identified by anonymized participant number (e.g., P12). The complete set of participant responses is provided as supplementary data in S1 Table.

\subsection*{Participants}
We collected responses from $N=70$ CP users. This sample size of real-world users is on par with or greater than those reported in recent published CP research~\cite{park2019tunes,park2021social}.
Participants were recruited through various music-related groups on social media, listservs, and flyers. Participants ranged in age from 18 to 59 years ($M=24.2$, $SD=7.8$), 51\% were female, and all used Spotify to engage in CPs. All participants were from the United States or a territory thereof. Responses were given anonymously; upon completing the study, participants were redirected to a separate form to input their email address to be entered into a raffle with 10\% of winning a \$10 Amazon gift card. All responses were collected prior to the onset of COVID-19 shutdowns. 

All but one participant reported at least one useful or important aspect of CPs (Q1), and all but two reported lacking or desired aspects (Q2). 
For Q1, text responses ranged in length from 1 to 58 words ($M=13.9$, $SD=13.3$ words). For Q2, where participants wrote about both shortcomings and desired features, responses ranged from 2 to 218 words in length ($M=26.7$, $SD=29.5$ words). 

\section*{Results}\label{sec:results}
Our thematic analysis yielded the Codebook of Critical CP Factors, which consists of eight high-level aspects. The eight high-level aspects of the codebook, and their sub-categories where applicable, are summarized in Table~\ref{tab:Codebook}. 

\begin{table}
    \caption{Codebook of Critical CP Factors, derived from thematic analysis of free-text responses delivered by CP users regarding which aspects of CPs are most useful and important, and which are desired or lacking.}
    \centering
    \begin{tabular}{p{.45\linewidth}p{.45\linewidth}}
    \toprule
    \begin{tabular}[t]{p{.95\linewidth}}
    \textbf{1. Content}. Pertaining to characteristics or quality of the songs in the CP, e.g., the \textbf{variety}, \textbf{diversity}, or \textbf{newness} of songs; the musical \textbf{theme} of the CP; songs being \textbf{reflective of all tastes}; or the need for \textbf{quality control} regarding the songs in the CP. \\ \\
    \textbf{2. Discovery}. Pertaining to inwardly directed acts of \textbf{learning} or \textbf{receiving information} in the course of engaging with a CP. Examples: Discovery of \textbf{music}; learning about \textbf{collaborators' musical tastes}; discovering \textbf{new collaborators} or \textbf{new CPs} with which to engage. \\ \\
    \textbf{3. Social}. Pertaining to outwardly directed acts of \textbf{sharing} or \textbf{recommending} music, and to a CP facilitating or signifying \textbf{social connections} among collaborators. Social codes: \\
    \small{3.1. Share} \\
    \small{3.2. Recommend} \\
    \small{3.3. Connect} \\
    \small{3.4. Heart behind item} \\
    \\
    
    \textbf{4. Platform}. Pertaining to characteristics of the streaming platform: \textbf{Ease of use}, \textbf{presentation of content}, and \textbf{access} to the platform and its content. Platform codes are as follows:  \\
    \small{4.1. User experience} \\
    \small{4.2. Content organization} \\
    \small{4.3. Access} \\
     \\
    \end{tabular}
    &
    \begin{tabular}[t]{p{.95\linewidth}}
    \textbf{5. Consumption}. Pertaining to how users listen to the CP, whether \textbf{alone} or \textbf{with others}, \textbf{synchronously} or \textbf{asynchronously}. Pertaining to the ordering of songs when a CP is played (i.e., \textbf{shuffling}). \\ \\
    
    \textbf{6. Editing}. Pertaining to the act of editing the CP (alone or with \textbf{multiple editors}), \textbf{control} of others' ability to edit the CP, or accessing a \textbf{historical record} of edits. The following four aspects of editing were coded: \\
    \small{6.1. Editing artifact} \\
    \small{6.2. Multiple editors} \\
    \small{6.3. Control settings} \\
    \small{6.4. Record} \\
    \\
    \textbf{7. Visibility}. Pertaining to the extent to which users' \textbf{edits}, \textbf{consumption patterns}, and \textbf{sentiments} are made visible to themselves or to others. The three aspects of CP visibility are as follows: \\
    \small{7.1. Of edits} \\
    \small{7.2. Of consumption} \\
    \small{7.3. Of communication} \\
    \\
    \textbf{8. Engagement}. Pertaining to the extent and nature of user engagement with a CP, e.g., \textbf{longevity} of a CP collaboration, \textbf{awareness} of CPs in general or a particular CP, the \textbf{intrinsic process} of using a CP, (inter-)\linebreak personal \textbf{feelings} that help or hinder the collaboration. \\
    \end{tabular}
    \tabularnewline
    \bottomrule
    \end{tabular}
    \label{tab:Codebook}
\end{table}

Across the collection of responses, all eight aspects were referenced with regard to both their usefulness and importance (Q1) as well as their lacking or desired aspects (Q2). In addition, some responses referenced codebook categories conceptually, others mentioned specific proposed features that could improve them, and some responses were both broad and specific. Proportions of mentions for each aspect of the codebook are summarized in Table~\ref{tab:CodePercentSummary}.

\begin{table}
    \caption{Summary of the percentages of free-text responses mentioning each of the eight top-level aspects in the Codebook of Critical CP Factors. Bold text in each column indicates the top two categories for reported useful/important and lacking/desired aspects.}
    \centering
    \begin{tabular}{lcc}
    \toprule
    CP Aspect & Useful/Important & Lacking/Desired \\
    \midrule
    1. Content & 26\% & 16\% \\
    2. Discovery & 13\% & 3\% \\
    3. Social & 13\% & 13\% \\
    4. Platform & \textbf{31\%} & \textbf{44\%} \\
    5. Consumption & 9\% & 11\% \\
    6. Editing & \textbf{47\%} & 16\% \\
    7. Visibility & 20\% & \textbf{31\%} \\
    8. Engagement & 6\% & 21\% \\
    \bottomrule
    \end{tabular}
    \label{tab:CodePercentSummary}
\end{table}

\begin{figure}[ht]
    \centering
    \includegraphics[width=1\columnwidth]{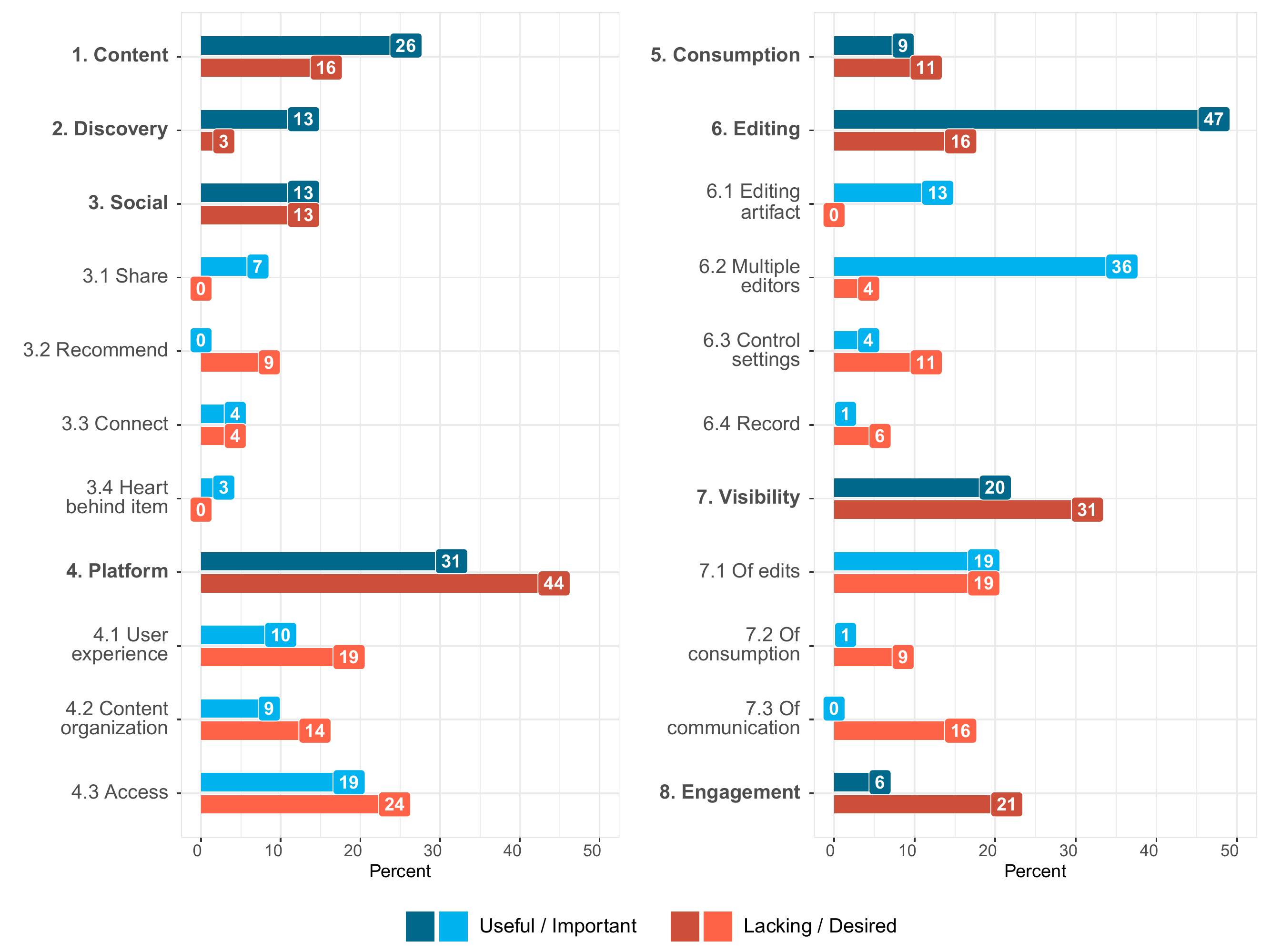}
    \caption{Percentages of free-text responses mentioning each of the aspects in the Codebook of Critical CP Factors that were coded (Table~\ref{tab:Codebook}). Bold text indicates the top-level factors.}
    \label{fig:CriticalFactors}
\end{figure}

\subsection*{Content}\label{sec:content}
Responses in the first theme,  Content, were related to characteristics or quality of the songs in the CP. This category aligns with the Practical purpose of the CP Framework~\cite{park2019tunes} that implicates content characteristics of CPs. As shown in Figure~\ref{fig:CriticalFactors}, the majority of responses in this category highlighted useful and important aspects of Content (26\%) as opposed to lacking or desired aspects (16\%). Users appreciated the diversity or \textit{``variety''} (P12, P33) in the CP and the way the music of a CP represented \textit{``selections that reflect the authentic collaboration of those who put the playlist together''} (P22) that \textit{``can bring together many different tastes''} (P47). 
Others stated that the CPs enabled playlists to form with \textit{``some amount of coherency in the music''} (P11) and \textit{``continuity''} (P27), which could imply a consistent \textit{``theme''} (P30, P54) of the music, or \textit{``music mixes''} (P25) for specific occasions such as \textit{``when we are together or at a party''} (P44).

Lacking or desired elements to better support the Content aspect of CPs included the need for more \textit{``quality control''} (P52) of the music added, which could become an issue when \textit{``people get carried away and add bad music and take over a playlist''} (P51). This issue---and, more broadly, CP co-curation involving more song additions than deletions---meant that CPs could become \textit{``extremely long''} (P15) and therefore difficult to manage: \textit{``They get long very quickly and then when they're too long it's annoying to shuffle through them because there's a high chance you'll come across a song you don't like''} (P38). 
P38 additionally suggests this could be \textit{``a shortcoming on the collaborators' efforts to make a NEW playlist or delete songs''}. As solutions,  P51 proposes to \textit{``limit the number of songs a collaborator can add''}, and P15 suggests \textit{``a feature that could divide the playlist into sub-blocks that could be moved around''} to manage large quantities of music.
Overall, multiple mentions of similar sentiments highlight the difficulty of maintaining a shared playlist consisting of content that everyone enjoys.

Interestingly, other shortcomings contrasted directly with advantages underscored in positive responses---e.g., that a CP had \textit{``no diversity within song selection} (P27) or was \textit{``too focused on popular music''} (P46). These and other reports---such as \textit{``it is hard to get music that everyone likes''} (P30) and \textit{``not everyone has the same or similar music tastes''} (P70)---reflect the variety of experiences that CP users may encounter.

\subsection*{Discovery}\label{sec:discovery}
Responses in the Discovery aspect pertained to learning or receiving information about music, playlists, and even collaborators. This category is well aligned with the Cognitive category of the CP Framework~\cite{park2019tunes}. Regarding useful and important Discovery attributes (13\% of responses), CPs were reported to aid users in \textit{``discovering new songs I would not find otherwise''} (P65) and \textit{``new music and artists I love''} (P53). For some, CPs also acted as a portal for receiving an \textit{``influx of new and fresh but still pertinent music''} (P19) as well as \textit{``recommendations based on what I listen to''} (P62). Many users also expanded upon learning about collaborators' music, underscoring the importance of CPs in \textit{``[seeing]\ who adds which songs so I can learn what music my friends like''} (P15) and learning \textit{``what my family is listening to''} (P61). Moreover, P12 stated that \textit{``I think having the ability to [...]\ discover new songs that a person you know (such as my roommate) listens to is cool''} and P27 further notes that CPs help them better understand collaborators as people: \textit{``What inspires them, what moves them, what do they believe in?''} 

Discovery was the theme with the fewest mentions of lacking/desired aspects (only 3\% of responses): P10 expressed difficulty in discovering CPs (\textit{``it's kind of hard to find good collaborative playlists''}), while P14 desired recommendations for potential collaborators (\textit{``I would love it if the application suggested WHO to create a collaborative playlist with out of my Facebook friends based on similar music tastes''}).

\subsection*{Social}\label{sec:social}
The Social aspect of CPs and their usage included mentions of sharing and recommending playlists, using a playlist as a venue for sharing music, or connecting with others and expressing sentiments through CP curation actions. These attributes align with the Social purpose of the CP Framework~\cite{park2019tunes}. In the present data,  mentions of useful or important Social aspects were on par with desired or lacking aspects (13\% of responses each).

With regard to sharing, users stated that CPs enabled them to \textit{``share music''} (P12, P44) and express \textit{``shared interests between friends''} (P41). No users mentioned a lack of sharing in CPs. For the more specific aspect of recommending, however, users mostly mentioned a desire for the platform to help them recommend music to others through CPs. This ranged from \textit{``recommending songs from each other's playlists''} (P43) to \textit{``songs similar to the ones in your playlist''} (P45), because there are \textit{``so many songs in your playlist that it is hard to remember which song [...]\ would fit the playlist''} (P63). Users also mentioned a desire for recommendations based on songs already in the CP: \textit{``Similar song suggestions collaborators can consider adding''} (P10). 

Social connections among collaborators were also noted. On the positive side, users mentioned that encountering others' contributions to a CP can reflect \textit{``relatability''} (P46) and \textit{``lots of heart in what songs that person puts on the playlist''} (P27), and make them \textit{``think of the person that you made it with''} (P14). Yet users also pointed out that currently the social aspect was lacking (P48) and, as noted in Discovery above, that the platform could make better use of users' social connections on other platforms such as Facebook to suggest potential CP collaborators (P14).

\subsection*{Platform}\label{sec:platform}
Responses in the Platform aspect pertained to the user interface and experience, as well as access to the streaming platform. This category was most mentioned overall, with 31\% of responses noting useful or important aspects, and 44\% of responses noting aspects that were desired or lacking. 
For user experience, responses referenced the ease of usage and contributions for CPs, e.g., \textit{``the ability to easily drag songs from my playlists and add to another''} (P15) and \textit{``clean easy to use interface''} (P49). However, users also mentioned difficulty and confusion around starting or locating CPs: \textit{``Spotify doesn't have a very easy interface for creating collaborative playlists''} (P8) and \textit{``it was very difficult for my friend and me to figure out how to create and share a collaborative playlist on Spotify''} (P53). Some even stated that the CP itself was not easy to use: \textit{``They're just hard to use, even when everyone uses that platform''} (P40).

With regard to content organization, users highlighted basic playlist functions not unique to CPs, such as \textit{``being able to manually sort/organize/order the playlist''} (P7). But users also desired additional capabilities to organize content, whether permanently or temporarily. Two common ways in which users wanted to organize CP contents were by filtering and sorting---for instance by genre (P5), contributor (P7), listening context (P15), newness (P46), or to \textit{``distinguish recently added/temporarily added songs from more fixed songs on the playlist''} (P26).
Some users also hinted that automation and AI could support some of these reordering functionalities (e.g., \emph{``filter out newer music''} (P46), \emph{``push [top-rated songs] to the top of the list''} (P2)). 

Finally, with regard to access, users found helpful the ease of contributing or navigating to CPs from their personal music and artist pages (P15, P24), offline access (P32), and access to the variety of music afforded by Spotify (P33). P37 particularly appreciated the varying levels of access to playlists, including CPs: \textit{``the public vs.\ private, collaborative vs.\ view-only settings are nice''}. However, within the platform, many users felt that sharing access was confusing, stating that CPs \textit{``can be confusing to share / set correct privacy settings''} (P31). Moreover, platform differences, leading to restrictions in access to CP usage and sharing, were cited as a problem for many users (e.g., \textit{``if everyone used the same music platform''} (P11), \textit{``it is hard to make playlists when your friends use different music platforms than you'' (P21)}). Another lacking aspect of access was the inability to add songs to the CP from outside of Spotify, and some users desired the \textit{``ability to add songs from the radio or out and about''} (P50).

\subsection*{Consumption}\label{sec:consumption}
Responses coded under the Consumption aspect of CPs pertained to how users consumed (i.e., listened to) the music in the CP, whether synchronously or asynchronously. 
Useful or important aspects of Consumption were noted in 9\% of responses. The ability to consume the CP in different contexts was seen as important (e.g., \textit{``when we are on our own or [...]\ when we are together or at a party''} (P44)). Users also valued the ability to access and play the CP on their own: \textit{``I like being able to go at it at your own pace. The best feeling is when you do not add something for a while and then you check the playlist and 20 songs were added and you can just start listening to them all''} (P14). Other responses highlighted Consumption functions of playlists more broadly, not specific to CPs: \textit{``Being able to listen any time''} (P31), \textit{``being able to shuffle the playlist or play specific songs''} (P7).

In lacking or desired Consumption-related CP attributes (11\% of responses), a number of users expressed a desire for synchronous co-consumption: \textit{``Being able to listen at the same time (synchronized) on separate devices in separate locations''} (P31). Users also desired what is  referred to as \textit{``queues''} (P1, P55)---the ability for groups \textit{``to listen to songs with everybody contributing in real time''} (P1). 

Additional features to support synchronous consumption with collaborators were also requested: \textit{``A DJ mode, where you can create transitions between songs''} (P17), a function to \textit{``divvy whose song gets played when''} (P34), a function to \textit{``\thinspace`hide' songs that you personally don't [want]\ to hear when you shuffle the playlist (without deleting them from the entire playlist!)''} (P38), and \textit{``a function where I could play songs that were added to the playlist by a specific person/people, or maybe more generally, a function that let me play only songs contributed by other people and no songs I contributed myself''} (P7). 

\subsection*{Editing}\label{sec:editing}
Editing refers to acts of adding, deleting, or reordering songs in the CP; control settings of editing rights; and historical record of edits and versions of the CP. Editing was notably the theme with the greatest number of mentions around being useful or important for users (47\% of responses), while lacking or desired aspects of Editing were mentioned in 16\% of responses. Multiple users mentioned the general ability to edit the CP, for example \textit{``the ability to dynamically edit the playlist over time (e.g., number position of song, deleting a song, etc.)''} (P5). The ease of \textit{``being able to contribute any songs''} (P31) was also mentioned as important: \textit{``The ability to easily drag songs from my playlists and add to another, preferably in volume [and]\ the ability to rearrange and delete songs on the collaborative playlist is important too''} (P15). Synchronous editing was also noted as useful: \textit{``Can both update the playlist at the same time''} (P32). 

More frequently appreciated was the \textit{``collaborative aspect''} (P4) of having multiple CP editors: \textit{``Anyone being able to contribute''} (P23) and \textit{``everyone who has access can add their songs''} (P3), thereby enabling \textit{``all users able to change playlists equally''} (P8). In short, the \textit{``simultaneous cloud-based collaboration''} (P24) was stated most frequently as being useful. Yet while distributed, synchronous editing is already offered on Spotify, some responses suggested that not all users were aware this was the case; for instance, P8 felt that the CP experience would be enhanced \textit{``if we could both work on it from our respective devices at any time''}.

At the time of writing, all collaborators on a Spotify CP have equal edit rights to the playlist. A number of users viewed this as being useful: \textit{``Anyone''} (P3) who has access to the CP can change the playlist \textit{``equally''} (P8). Other users, however, desired more fine-tuned settings such as \textit{``an optional `admin' role for the playlist and approval of adding/deleting songs, as well as a fully equal-party mode instead of that option''} (P2), or for only \textit{``certain people [to be]\ able to add songs''} (P37). These control settings, also called \textit{``editing permissions''} (P19), were desired in response to pain points in adding to and deleting songs from the CP: \textit{``Hard to manage what gets [added]''} (P51) and \textit{``someone can just delete your song and you'd never know who it was''} (P7). For such issues, users proposed specific solutions such as putting \textit{``limits to how many songs people can add [and]\ 2-person agreed deletion of songs''} (P4) or allowing \textit{``control of lead collaborators and having specific options for others''} (P64).

The ability to access a historical record of edits was also noted as lacking: \textit{``I can't see the history of what was removed/added, in case I want to undo an action''} (P2). P7 lamented that \textit{``there's not really a history function and [...]\ someone can just delete your song and [you may]\ not even know it was ever in the playlist to begin with''}. Some also wished for music to be experimented with off-record, akin to a staging area or draft of the playlist: \textit{``Being able to distinguish recently added/temporarily added songs from more fixed songs on the playlist [because]\ you have to create another playlist to experiment''} (P26). Overall, these types of co-editing capabilities are summarized in one user's desire for CP functionalities to be more like Google Docs: \textit{``Group editing techniques‚ similar to Google Docs would enhance such collaboration [as]\ it can be difficult to make updates and collaborate through the same interface''} (P22).

\subsection*{Visibility}\label{sec:visibility}
Responses reflecting Visibility pertained to CP edits, consumption, and communication being shown to users. Useful and important aspects of Visibility were mentioned in 20\% of responses, while lacking or desired aspects were mentioned in 31\% of responses.
Multiple users highlighted the importance of edits being made visible, for example in \textit{``seeing which users added which songs''} (P2) and \textit{``knowing who put what songs in the playlist''} (P7)---\textit{``(and to a lesser degree, when)''} (P24). Viewing these edits, especially song additions, enables users to \textit{``learn what music my friends like''} (P15). 
While at the time of data collection, songs in a CP are labeled with the handle of the user who added each one, some expressed this as a desired feature: \textit{``I think it would be cool to see who added what although I use Spotify and that's not currently a feature''} (P67).
Moreover, multiple users expressed a desire to view a more extensive history of edits, e.g., \textit{``in case I want to undo an action''} (P2). Numerous users (14\%) desired more editing visibility than just history. For instance, multiple users wanted to receive notifications when the CP was edited, for instance as a way to know of edits without having to check the CP themselves: \textit{``Having a way to opt-in to notifications of when people add songs to the playlist would be cool so I just know when people add new songs (and what they are) instead of not realizing that they did or having to check regularly''} (P7).

While visibility of edits were mentioned equally in terms of useful/important and lacking/desired aspects (19\% of responses each), the other aspects of visibility---i.e., consumption and communication---leaned more toward lacking or desired in current CP usage. For visibility of consumption, one user did feel that music included in the CPs could serve as a proxy for consumption, stating that CPs help them understand \textit{``what my family is listening to''} (P61). But others who felt that visibility of consumption was lacking or desired stated that \textit{``it's sort of hard to see if other people are actually listening to the playlist or not''} (P7) or \textit{``if I've never listened to the song''} (P30), highlighting a lack of awareness of both others' and one's own CP music consumption. This desire for visibility of consumption also translated to requests for features providing \textit{``analytics [to see]\ when the collaborator listens''} (P34).

Visibility of communication was mentioned solely as a lacking or desired aspect of CPs. Many users desired to \textit{``discuss the playlist''} (P15) and the songs: \textit{``Maintaining an open, active conversation regarding new songs that should be added and old ones that should perhaps be removed''} (P27). Users also wished to \textit{``provide feedback to each other on the songs that each collaborator adds''} (P37), which could facilitate collective decision-making such as \textit{``2-person agreed deletion of songs''} (P4). 16\% of participants mentioned specific features promoting visible communication, e.g., for \textit{``individual collaborators to rate/comment on songs in the playlist, whether they like it or not, and other things''} (P7) and to provide commentary on \textit{``why they like it [would be]\ nice''} (P28). In addition to \textit{``rate''} (P2, P7, P15, P52) and \textit{``comment''} (or \textit{``chat''}) (P7, P15, P28, P39, P59) features, users also mentioned a wish for \textit{``notes attached to songs''} (P24) and \textit{``likes''} (P39).

\subsection*{Engagement}\label{sec:engagement}
Engagement, the final category in the codebook, pertains to mentions of longevity, awareness, and intrinsic process of CPs, as well as feelings surrounding their use. 
Only 6\% of responses mentioned useful aspects of Engagement, e.g., the enjoyment of creating, curating, and consuming CPs. One user also noted the \textit{``humor with which you can approach the process''} (P19).

Meanwhile, Engagement accounted for the third-most lacking mentions among the codebook categories at 21\% of mentions. Hinting at CPs falling short of their desired longevity or sustained engagement over time, one user stated that \textit{``they are only used for a short period of time''} (P18). Another lamented that collaborations around CPs were not as engaging as they would like: \textit{``I don't feel as engaged to work on them. [CPs]\ have been created with the intention of being collaborative, but no one else ends up adding or curating them''} (P35).

Overall, responses in this category highlighted a desire for greater engagement with CPs, coupled with a lack of awareness of current engagement.
Many users wanted to know whether others were engaging, whether through edits or consumption. For one, P35 felt that visible edits, as operationalized by notifications, would motivate greater engagement: \textit{``I think that notifications indicating when things are adjusted and changed would help to encourage more interaction''}. Such notifications may even \textit{``[get]\ a group to regularly contribute''} (P68). Others wished for notifications in the \textit{absence} of engagement, as CPs \textit{``often get forgotten/not added to after a while''} (P50). Here, users suggested that CPs \textit{``should be subtly promoted on the streaming platforms''} (P50) and showed desire for \textit{``a notification from the playlist platform to check out the collaborative playlist when it starts becoming inactive''} (P41). One user even noted that \textit{``not everyone is aware of collaborative playlists''} (P62), suggesting that awareness of and engagement with CPs could be improved on even a broader level.

In terms of the intrinsic process of CP usage, one user suggested \textit{``[making]\ it more `fun'\thinspace''} and providing more of a \textit{``\thinspace`social platform' appeal''} (P48). Finally, with regard to feelings---which could be personal or inter-personal---participants mentioned anxieties around sharing music (e.g., \emph{``afraid to add songs that others may not like''} (P9)), challenges related to disparate musical tastes (e.g., \emph{``hard to get music that everyone likes''} (P30)), and unwanted behaviors among collaborators (e.g., getting \emph{``carried away''} and \textit{``tak[ing]\ over a playlist''} (P51)).

\section*{Discussion}\label{sec:discussion}
We surveyed users of collaborative playlists and conducted thematic analysis of open-ended text responses to better understand what aspects of CPs and their usage they find useful, important, desired, and lacking. Our analyses uncovered eight high-level aspects of CPs and their usage: Content, Discovery, Social, Platform, Consumption, Editing, Visibility, and Engagement (Table~\ref{tab:Codebook}). 

At the time of data collection, CPs were essentially ``personal-plus'' artifacts---that is, playlists with co-editing functionality and indication of who contributed each song in the playlist and when. Unsurprisingly, participants confirmed that these basic functionalities are useful aspects of the CP experience. However, results also suggest more to be done to support user needs around CPs. Our results and prior work collectively point to numerous lacking or desired features in each of the eight codebook categories, highlighting opportunities for engaging users further in CPs. From these insights, we offer design implications for CP platforms.

\subsection*{Design of CP functionalities ought to be informed by users' purposes for engaging in CPs}\label{sec:discussCPFramework}
Three of the eight aspects of CP usage that emerged from participant responses (Content, Discovery, and Social) directly aligned with the Practical, Cognitive, and Social purposes of the CP Framework~\cite{park2019tunes}, respectively. 
Each of these aspects was mentioned with varying extents of positive and negative implications, suggesting that purposes for which people engage in CPs are fulfilled in practice, though there is room for streaming platforms to improve user experiences to serve CP purposes further.

These categories echo reports social music prototype research, underscoring our findings that social curation and consumption empower users to achieve music goals. Aspects of Content, the most-mentioned CP Framework category among our responses, are noted in previous work. For instance, Lehtiniemi et al.\ reported that users ``thought more carefully about the content they added to the service since they were more socially aware of other users'', creating ``positive pressure for content creation''~\cite{lehtiniemi2017socially}. But prototype users have also reported challenges of too much music or bad music~\cite{hakansson2007pushMusic,liu2008social}---the latter of which could compel users to abandon the social platform altogether if too many bad songs were presented in succession~\cite{liu2008social}.

Responses positively implicating the Discovery aspect of CPs also align with past research. With Jukola, for example ``one of the key values of the system was learning about new music''~\cite{ohara2004jukola}; this was also reported by users of tunA~\cite{bassoli2006tunaSocialisingMusic} and Push!Music~\cite{hakansson2007pushMusic}. In the present study, users reported multifaceted perspectives of Discovery, reminiscent of the potential of social prototypes to help users ``discover new music through their friends, and [make]\ new discoveries about the friends as well''~\cite{liu2008social}. Finally, discovery of new collaborators---a wish expressed in our current data---was also reported in a number of prototype studies~\cite{chao2005adaptiveRadio,bassoli2006tunaSocialisingMusic,lehtiniemi2013evaluating}; this can then bring about Social benefits, e.g., ``creat[ing]\ new meaningful social connections [...]\ having a communication channel to start a `virtual conversation'\thinspace''~\cite{bassoli2006tunaSocialisingMusic}.

Most responses implicating Content and Discovery were positive, suggesting that current CP design serves these purposes. However, mentions of the Social aspect were mixed in sentiment in our current sample. Similar to reports that users enjoyed sharing music through such platforms~\cite{hakansson2007pushMusic} and that doing so can strengthen relationships~\cite{liu2008social}, the heart behind songs, while mentioned rarely, evinces the dual social aspects of sharing music and bonding with others. In addition, the potential for CPs to heighten social outcomes through collective reminiscence has been noted, e.g., in posted historical records of ``winning'' Jukola songs~\cite{ohara2004jukola} and synchronous listening while engaging over social media~\cite{lee2020youKnowYoure}. However, similar to present findings, past studies report that music contributions alone do not serve all of users' social needs (e.g., users of Social Playlist were ``not able to interpret the states of others by listening to the music itself''~\cite{liu2008social}). Given that users engage in CPs for Social and Practical purposes equally~\cite{park2019tunes}, features to fulfill Social aspects of CPs may also greatly enhance users' experiences.

\emph{Implication.} As current CP functionalities provide the bare minimum for collaboration, opportunities exist for CPs to better serve users in aspects reflecting CP Framework purposes, particularly Practical and Social purposes. We suggest that consideration of platform features be guided by users' purposes for engaging with CPs, with features evaluated against how well they meet the critical factors identified from this work. For example, current designs neither consider social awareness aspects that can lead to more successful CPs nor provide users avenues to reminisce together. Yet we see evidence that these functionalities can help users realize their Social and Practical purposes for CPs. For the Social purpose in particular, many desired more support in recommending music to others through the CP. While Spotify does provide a list of song recommendations below a CP, either people did not know about it (suggesting that settings for enabling this and the interface for reaching the setting are unclear) or felt the song recommendations given did not adequately reflect songs they wanted to add. This suggests ways in which the current design can be improved to better serve Social needs.
Thus we urge platform designers to ask questions such as: Does the feature aid Discovery? Does it facilitate greater Social connection? Does it lead to better Content or context? Expanding upon current functionalities can further clarify the value proposition of the CP platform, as the relationship between features and the purposes for which they are designed becomes more clear.

\subsection*{CP users seek improved platform access, UX, and content organization}
Platform aspects, investigated in previous literature, reflect the aspects that arose in our thematic analysis. While users provided many positive responses, all Platform sub-topics suggested much room for improvement. The most frequently occurring aspect was that of platform access. Issues regarding a lack of access without a login was stated most in the context of CPs. Many participants lamented that their friends all use different music services, and thus could not share music. This is a known issue with other collaborative platforms~\cite{posner1992howPeopleWrite}; on music streaming platforms and particularly for CPs, it points to platforms' support of individual music consumption while hindering social musical connections. 
A potential design solution, whereby CPs ``[make]\ it easier to curate a playlist across many different platforms'' (P40), could enable users to curate music together across platforms. Or, an open-access service that does not require authentication or a user ID (e.g., YouTube) could be introduced; this could also free users from the social implications of platform advocacy~\cite{park2021social}.

User experience of CPs was mixed, with some saying it was easy to navigate and add songs and others stating that even starting a CP was difficult. This adds to the ``hurdles and the friction'' associated with CP formation noted by Park \& Kaneshiro~\cite{park2021social}.
Moreover, several user responses suggested a lack of awareness of some existing CP functionalities---e.g., platform-recommended songs to add to the CP, version history, the ability for collaborators to edit from separate devices, and seeing who added which song.

Participants' desire for more tools to organize content reflects a common occurrence in CPs, which P15 stated aptly: \textit{``Sometimes the playlists are extremely long''}. This is due in part to collaborators \textit{``not deleting songs''} (P27), which may reflect hesitation and discomfort, noted by Park \& Lee, arising from perceived ownership of songs in CPs~\cite{park2021lost}. P38's desired functionality to \textit{``\thinspace`hide' songs [...]\ without deleting them from the entire playlist''} is one solution. Certainly the filter and sort functions noted by multiple users are one way of addressing this problem of ``long'' playlists. In addition, as P26 noted, separation of songs based on an articulated and agreed-upon criteria, and having a designated space in the interface, may be helpful. These features would also prove to be useful in managing personal playlists, which are also known to have sizing issues~\cite{redditspotify1,redditspotify2}. With regard to difficulties of deleting songs due to social norms and pressures, enabling a ``2-person agreed deletion'' (P4) functionality is one option. As noted in prior CP work~\cite{park2021lost}, enabling comfort felt in deleting songs in the CPs as well as contributions by other collaborators (e.g., by providing communication mechanisms) can mitigate this problem and lead to more positive experiences for the users navigating the CPs, which most likely contains more diversity of music that requires breaking down into parts that are more manageable.

\emph{Implication.} Features promoting better access, user experience, and organization of content (e.g., ``filter'' and ``sort'' functions available on spreadsheets) are common in other collaborative platforms (e.g., Google Sheets) and, if implemented for CPs, could improve the user experience. Some Platform issues, e.g., platform access, are more difficult to address than others. However, given the rising importance of technology in providing and maintaining social connections, platforms ought to be more mindful of the longer-lasting impacts to users and their interactions through their designs. Just as iPhone and Android users can communicate despite different operating systems, users of different music streaming platforms ought to be able to enjoy music together; applying a similar design will serve platforms well. Furthermore, the ability to synchronize CPs across platforms, such that content changes are reflected everywhere, could increase user engagement with a CP, in turn engendering higher engagement with the streaming platform. While we are cognizant that the collaborative experience on CPs is distinct from other virtual collaborations, especially as engagement and enjoyment involve both creation and consumption of the CP~\cite{park2021social}, users' responses suggest that implementing some of the lessons learned from other platforms may help in clarifying the initiation process of CPs and managing CP content and interactions. These considerations are particularly helpful for Spotify, which all our participants used to interact with CPs, as this elucidates concrete ways for the platform to enhance overall user experience.

\subsection*{CP users seek differentiated controls in social music curation}
Similar to Platform aspects, CP control settings noted in Editing could also benefit from insights from other collaborative platforms. Perhaps this is what one user was alluding to by stating \textit{``Group editing techniques‚ similar to Google Docs‚ would enhance such collaboration''} (P22).

Control is an important factor in personal playlist curation~\cite{morris2015controlCurationMusical,millecamp2018,hagen2015playlist} and for music curated for social contexts, such as parties and road trips~\cite{cunningham2014socialMusicIn,cunningham2009exploringSocialMusic}. Controlling others' access and actions has been highlighted in admin-type functionalities in music prototypes~\cite{hakansson2007pushMusic, lehtiniemi2013evaluating}. Many responses in the present study similarly expressed a desire for greater control in CPs. 
However, past research warns that not all users may want this type of control: With MoodPic, ``some participants were requesting ways to control the content of some specific playlists, where other playlists could remain open for additions [...]\ however, some did not want only the creator of the list to have all the control, but instead sharing the control to all of the users of the collective list''~\cite{lehtiniemi2013evaluating}.
Past social music systems have discussed attenuated forms of control over music content as well, exerted through forms of group influence such as ratings~\cite{liu2008social}, voting~\cite{ohara2004jukola}, or specified preferences~\cite{mccarthy1998musicfx}.
These types of functionalities appeared in our data sample, as were proposals to use filtering, sorting, and skipping to personalize the consumption of a shared playlist. Features that return control back to individual users within a group setting have the potential to decrease user anxiety (because they can perform actions without changing the public document)---an issue that has been noted previously~\cite{lehtiniemi2017socially,kirk2016understanding}---and increase engagement (e.g., if a user is especially interested in newly added songs or songs contributed by a specific collaborator). However, personalized sub-playlists could also lead to users creating personal playlists inside the CP, thereby interacting less with the shared playlist as a whole.

\emph{Implication.} Designing control elements---whether of collaborators' actions on the playlist, individual consumption, or both---is necessary to ensure satisfactory CP experiences. Contrary to the notion of control itself, these features can foster a sense of democratic collaboration and give users free rein to interact with the CP, bringing greater engagement and comfort to the collaboration. But they must be considered alongside potential tradeoffs between control versus active engagement~\cite{anbuhl2018social}. As explored in prior work~\cite{park2021social}, further exploration of specific features already implemented in other collaborative platforms can also help improve upon the current CP experience.

\subsection*{CP users want to better understand their collaborators}\label{sec:discussVisibility}
Visibility of others' interactions is known to be essential for successful collaborations on online platforms~\cite{erickson2000social}. One way in which CPs differ from other online collaborations is that the end product---that is, the playlist---may involve a greater degree of repeat consumption (e.g., for entertainment)~\cite{park2021social}. Consequently, and as suggested by the present data, collaborative interactions with a CP should consider Visibility not only of editing but also of consumption. 
Visibility of communication, too, could indirectly enable users to know whether collaborators have added to or listened to a CP, and provide insight into their feelings about a given track.

The desire for greater visibility expressed both here and in user interviews regarding successful CPs~\cite{park2021social} suggests that collaborators want to better understand one another. 
Multiple contributors acting upon the CP introduce the possibility of missing out on both the \textit{event} of modification (e.g., a song added or deleted between between visits) and what this modification \textit{signals} (e.g., why a collaborator added a song).
Users also reported that it is difficult to know whether or how collaborators enjoy or engage with songs in a CP. In other words, current ``personal-plus'' implementations do not adequately communicate information that users want.
The desire to access and contribute information---from how many times a collaborator has listened to a song, to a ``like'' or comment---highlights a current gap in communication between collaborators. 

\subsubsection*{Of edits and consumption (what others do)}
Wishing to access the version history of a CP shows a desire to understand not only a playlist's content but also its process, a sentiment reflected in past work~\cite{hakansson2007pushMusic}. The consequent ability to undo changes may also give collaborators more psychological safety or comfort to edit a CP. 
Users' perceptions of functions are also highly critical. For example, as mentioned previously, one response suggested a lack of awareness of an existing functionality of seeing ``who added what'' (P67). This perception is critical to the users' CP experience as it also influences how they interact through the CPs. Hence we need to first understand why such perceptions exist and then address them in designing features. While further investigation is needed to accurately explain why, we speculate that this particular user perception may arise because the platform's desktop application is showing usernames only, or the user interacts with CPs only through a mobile application that does not make this immediately visible.
Furthermore, having this knowledge and awareness of who made which edits in a CP---whether gained by accessing the playlist or by receiving a notification---can help users achieve a sufficient level of workspace awareness~\cite{gutwin1996workspaceAwareness,gutwin1998effects} and improve the collaboration.

Users also wanted to know how collaborators were consuming CP content. This has been highlighted in past research, and motivations can range from curiosity \cite{lehtiniemi2013evaluating,hakansson2007pushMusic} to wanting to better understand others' musical tastes \cite{kirk2016understanding}. Past studies have also reported that users feel a sense of accomplishment or motivation from knowing that others are listening to content that they contributed \cite{ohara2004jukola,liu2008social,lehtiniemi2013evaluating}. Visibility of consumption can also promote engagement, as ``information on how many users have listened to their playlist [...]\ was said to motivate playlist creation even more'' \cite{lehtiniemi2017socially}. We surmise that knowledge of CP consumption behaviors could also help users identify songs to add or delete. Finally, transparency has the potential to strengthen interpersonal dynamics~\cite{liu2008social,lehtiniemi2013evaluating}, and a better understanding collaborators' tastes (by knowing what they are doing) can diminish anxiety around adding songs~\cite{kirk2016understanding}, 

However, surfacing consumption behaviors may bring about less-desired outcomes as well. Self-consciousness over how others view one's music tastes (e.g., in terms of sharing and consumption) has been noted in past studies~\cite{voida2005listeningInPractices,kirk2016understanding,liu2008social}. Users have expressed contradictory wishes between wanting to know what others are doing, but not wanting their own actions to be tracked~\cite{anbuhl2018social}. Along these lines, the extent of transparency in CPs could fundamentally change how users listen. With Pocketsong, Kirk et al.\ reported that users changed their listening habits as a result of their listening record being visible, ``shap[ing]\ their listening not just for their consumption but explicitly for the consumption of others''~\cite{kirk2016understanding}. Therefore, care is needed to strike a balance between requests for transparency, and users' privacy needs and comfort levels.

\emph{Implication.} Users seek more information on how CPs are edited and consumed. We recommend that platform designers carefully consider how a feature will balance needs for both transparency and privacy, which will likely vary by activity and user. This might entail providing granularity in transparency settings, such as an option to keep some listening sessions private. Or, users could select the extent of personal data they share on a per-playlist basis, as in Beierle et al.~\cite{beierle2016privacyAwareSocial}---perhaps in a reciprocal fashion (where the visibility a user grants others is the visibility granted to that user). Learning from research on change awareness~\cite{tam2006framework} with mechanisms such as those for version control~\cite{kraut1992task} to increase granularity in visibility of editing would be helpful.
Such nuances in visibility will help to motivate collaborators toward greater engagement, psychological safety, and group dynamics in CPs, while addressing worries around privacy and impression management.

\subsubsection*{Of communication (what others think)}
The act of music sharing is a noted form of social interaction~\cite{kirk2016understanding}. Even so, the social music behavior of conversing around music is at risk of being lost in the digital age~\cite{brown2001music}. Our present results indicate that users seek additional communication features such as likes, ratings, and comments. These expressions are known to improve collaborations over online platforms~\cite{kou2014playing} and provide a sense of copresence on other music platforms such as SoundCloud~\cite{hubbles2017f}. A number of benefits are noted in social music prototypes as well: Additional means of communication facilitate richer interactions~\cite{kirk2016understanding} and music discovery~\cite{ohara2004jukola}; let users know whether others liked their song recommendations~\cite{hakansson2007pushMusic}; enable users to explain selections, potentially reducing misinterpretations~\cite{liu2008social}; and help to form or strengthen social relationships~\cite{bassoli2006tunaSocialisingMusic,liu2008social}.

On the other hand, prototype users have expressed concerns that comments would ``fill the whole service with irrelevant content''~\cite{lehtiniemi2017socially}, and that general communication (though not visibility) features would displace music as the central focus~\cite{lehtiniemi2013evaluating}. These results are corroborated by findings that more users would feel more comfortable about taking actions on CPs if given a channel to communicate~\cite{park2021lost}.
Finally, prototype users have reported a risk of feeling bad about contributed musical content if others respond to it negatively~\cite{lehtiniemi2013evaluating}.

\emph{Implication.} Enabling CP users to communicate and exchange sentiments---whether through comments, likes, ratings, or an alternate features altogether---can further enhance engagement with the playlist and the music platform. It will be imperative, however, that communication features do not take over the music. As with history and transparency, optimal design settings will likely differ according to the expression and the user. Enabling situated exchange with annotations~\cite{baecker1995user} to address the visibility of communication may a starting point.
What the platform should do with the sentiments also remains an open question. For instance, recent research has shown that people react differently to positive versus negative feedback in group playlist curation~\cite{bauer2020conformityBehavior}. Therefore, continued research will be needed to understand how platforms can best surface and act upon collaborators' expressed sentiments.

\subsection*{Users seek active engagement with CPs and look to system interventions to support their engagement}\label{sec:automation}

An active community is known to be critical to the success of social music platforms~\cite{lehtiniemi2017socially} and of collaborative content in general~\cite{olsson2009collectiveContentAs}. In our study, Engagement was largely a lacking or desired aspect of CP usage. For instance, users reported that they or their collaborators did not engage enough in the CP, even though there was a desire to do so. Some suggested that even if they were too busy to pay attention to CPs, they would like to. Others felt engaged, but felt that others were not matching their level of involvement. Park \& Lee suggest that issues of perceived ownership could also affect CP engagement~\cite{park2021lost}; hesitations around deleting and curating music in CPs could hinder users from further consuming them. This underscores participants \emph{wanting} themselves as well as others to engage in CPs. 
Regardless of whether this desire is due to collaborators themselves not being able to engage as actively as they would like, or misaligned expectations and wanting others to contribute more, it needs to be addressed. 

Engagement was an underlying, and perhaps also unifying, factor across all the eight codebook categories. 
First, lacking or desired features in other categories can cause engagement to suffer. For instance, bad or excess Content can cause people to disengage with a CP~\cite{liu2008social}, and inadequate Discovery of  CPs or collaborators can dampen CP engagement.
Just as the unprecedented scale of content accessible on streaming platforms can be overwhelming~\cite{cunningham2017exploring,cunningham2019interactingWithPersonal}, finding the right playlists and collaborators (e.g., given vast followership on social media) for co-curation is also of great import.
The Social aspect, too, is hindered when users do not participate because they do not know what music to share. Desired Platform characteristics are wide-ranging---from general usability to cross-platform capabilities---and prevent users from engaging with CPs to the desired extent. The two direct interactions with CPs---Editing and Consumption---combined with a desire for greater Visibility---of edits, consumption, and communication---also point to CP users' desire for higher engagement with the playlists. A lack of co-editing and control features common to other collaborative systems, as well as a lack of visibility as to what others are doing or thinking, inhibit both the procedural aspects of CP curation and the music-related conversations and interactions that are a critical piece of social music behaviors~\cite{cunningham2019interactingWithPersonal}. Above enabling users to better know and interact with CPs, platforms can intervene to improve user experience. Many users' desire for notifications, particularly when songs are added, hint at their untapped potential for Engagement. 

At the same time, successful engagement can positively feed back into other aspects of CP usage. For instance, an active community of collaborators can create ``positive pressure for content creation''~\cite{lehtiniemi2017socially}, while greater engagement with CPs can bring about more contributions, which would enhance Discovery. In turn, understanding others' tastes better through their contributions to the CP, or through consumption behaviors, comments, or likes, can encourage others to contribute more and could also facilitate positive Social aspects of CP usage. These in turn can further drive engagement, perpetuating the positive cycle or ``Snowball effect'' of successful collective content~\cite{olsson2009collectiveContentAs}. Active participation was also a predictor of CP success~\cite{park2021social}---i.e., active engagement would beget even greater CP engagement. Finally, heightened engagement would provide rich information to the platform, whether implicit or explicit, which could inform further development of social music capabilities.

It is also important to note that some aspects of CP usage call for more urgent platform development than others. In particular, Platform features, control settings for Editing, Visibility of consumption and communication, and Engagement are all disproportionately reported in relation to lacking or desired features, while other aspects such as Discovery and Social, already better serve users' needs.
Since data collection, some aspects of the codebook have been addressed in commercial platforms.
For instance, Spotify has released Tastebuds, which aids Discovery as users can ``explore the music taste profiles of their friends''~\cite{constine2019spotify}. Updates to the CP enable users to more easily invite new collaborators and display avatars in front of each contribution~\cite{fortherecord2020how}.
Listening Together, an interactive globe displaying users streaming a song at the same time, was also released by Spotify in Spring of 2020 (\url{https://listeningtogether.atspotify.com/}),
as was Group Session, a shared queue for synchronous listening~\cite{perez2020spotify}. Both of these latter functionalities touch upon aspects of synchronous co-consumption of music---yet only 6\% of users in the present study expressed a need for any type of synchronous listening functionality (e.g., co-located or distanced), at least in the context of CPs. 

\emph{Implication.} 
Factors engendering successful engagement with CPs differ from those of personal playlists. That the burden of notifying others about a CP is currently placed on the user means missed opportunities for interactions, which in turn can bring about indifference and a lack of reciprocation in the CP. 
This is especially unfortunate given that a few bad collaborative experiences can discourage CP usage, even if users were initially optimistic and excited about it~\cite{liu2008social,vonSaucken2014userExperience}.
Thus, by supporting greater interaction with CPs, platforms can easily augment users' experiences. Determining appropriate methods of promoting Engagement---e.g., opt-in notifications of added song titles or connections to other songs in the CP---will be necessary. Yet features for heightened Engagement need not be completely reimagined; existing methods could be repurposed. For example, Olsson et al.\ outline the positive feedback cycle of engagement engendered by the ``Snowball effect''~\cite{olsson2009collectiveContentAs}; and AI and recommendation systems designed to help individual users navigate music~\cite{schedl2018current} could also be applied to finding potential CPs and collaborators~\cite{bassoli2006tunaSocialisingMusic}. In all, platform designers should critically examine ways in which current or planned functionalities facilitate or hinder user engagement with the CP.
Tying the current platform features with the type of factors can help platforms to deliberate and decide upon which features are important to include.

\section*{Future Works and Limitations}\label{sec:futureLimitations}
Our codebook offers a holistic framework detailing aspects of CPs that users find important. Future work can apply the codebook to study different types of CP users or use cases. For example, personas are known to provide a useful framing for investigating perceptions and behaviors of different types of users~\cite{anthony2009reputationAndReliability,lee2015understandingUsersPersonas}; if CP user personas also exist, they may implicate aspects of the codebook differently. Other studies could investigate impact of culture, collaborator group size and relation, and CP purpose. The codebook was derived from user responses reflecting CP usage prior to COVID-19, and it is likely that CP usage and user needs have changed since the onset of the pandemic. However, we argue that our findings---reflecting long-term usage of CPs---serve as a valid representation of user needs. Moreover, the codebook, reflecting pre-COVID usage, can serve as a useful basis for comparison in studies seeking to determine how musical practices---including usage of streaming platforms and social music curation---have changed in an era where virtual collaboration has become a necessity.

All participants engaged with CPs using Spotify, which likely influences the codebook categories, particularly Platform. Lacking or desired features reported by CP users are also necessarily speculative rather than experience based, and explicitly stated rather than implicitly observed. However, the extent to which our codebook categories align with past research on CPs, social music prototypes, and collaborative systems speak to their relevance for platform design implications. Future work should include direct observational and field studies to confirm whether proposed features bring enjoyment in practice or give rise to unforeseen challenges. Finally, new CP features and standalone applications have been introduced since these data were collected. 
We look forward to studying real-world outcomes of these new features, and hope that commercial platforms will additionally prioritize the most critical CP aspects identified here.

\section*{Conclusion}
Collaborative playlists (CPs) facilitate social music practices in the age of streaming.
Yet we lack a complete understanding of which aspects of CPs promote a positive user experience, and which are lacking. To address these questions, we conducted a survey in which real-world users reported on these facets of CPs and their usage.
Based on thematic analysis of free-text responses, we created the Codebook of Critical CP Factors, comprising eight aspects of CPs and their experience that users consider important: Content, Discovery, Social, Platform, Consumption, Editing, Visibility, and Engagement. Based on the responses and the extent to which they implicate each aspect, we could assess the level of importance these aspects have to users, as well as their specific benefits, obstacles, and suggested solutions. For instance, participants considered Editing (47\%) and Platform (31\%) aspects to be most useful, while Platform (44\%) was concurrently most lacking and desired, along with Visibility (31\%). These proportions of mentions cue us in to which aspects ought to be prioritized in order to enhance the CP experience from the user perspective.

Participants' reports of system needs reflect in-situ, long-term usage of a widely used commercial music streaming platform. Synthesizing these results alongside findings from prior works on social music prototypes and other collaborative platforms, we have derived design implications for CPs and their functionalities. The codebook and design implications together build a better understanding of CPs and contribute toward a more socially connected music experience, especially in this time of social and physical distancing. The user-centered design implications not only enable streaming platforms to prioritize certain aspects over others, but also have the potential to provide users with a more collaborative and enjoyable CP experience. In all, these contributions highlight the potential for streaming platforms to serve the needs not only of individuals, but of users consuming music together.

\section*{Supporting information}



\paragraph*{S1 Table.}
{\bf Free-text responses.} Anonymized free-text responses (with only participant numbers) from 70 real-world users of collaborative playlists, who reported on collaborative playlist features they found most important or useful (Q1) as well as observed shortcomings and desired features in collaborative playlist platforms (Q2).

\section*{Acknowledgments}
We thank Jonathan Berger, Arshiya Gupta, Camille, Noufi, and the Music Engagement Research Initiative at Stanford University's Center for Computer Research in Music and Acoustics for their invaluable feedback and support in this work.

\nolinenumbers


%
%
%

\end{document}